\documentclass[twocolumn,prl,aps,epsfig,graphics]{revtex4}
\usepackage{graphicx}

\def\gevfm3{$\frac{\mathrm{GeV}}{\mathrm{fm}^{3}}$}

\begin{document}

\title{Core-Corona Separation in Ultra-Relativistic
Heavy Ion Collisions}

\author{Klaus Werner}
\thanks{Electronic address: werner@subatech.in2p3.fr}
\affiliation{SUBATECH, University of Nantes -- IN2P3/CNRS-- EMN,  Nantes, France}
\begin{abstract}
Abstract: Simple geometrical considerations show that the collision
zone in high energy nuclear collisions may be divided into a central
part ({}``core''), with high energy densities, and a peripheral
part ({}``corona''), with smaller energy densities, more like in
pp or pA collisions. We present calculations which allow to separate
these two contributions, and which show that the corona contribution
is quite small (but not negligible) for central collisions, but gets
increasingly important with decreasing centrality. We will discuss
consequences concerning results obtained in heavy ion collisions at
the Relativistic Heavy Ion Collider (RHIC) and the Super Proton Synchrotron
(SPS).
\end{abstract}

\maketitle

Nuclear collisions at the Relativistic Heavy Ion Collider (RHIC) are
believed to provide sufficiently high energy densities to create a
thermalized quark-gluon fireball which expands by developing a strong
collective radial flow \cite{brahms-white,phenix-white,phobos-white,star-white}.
However, not all produced hadrons participate in this collective expansion:
the peripheral nucleons of either nucleus essentially perform independent
pp or pA-like interactions, with a very different particle production
compared to the high density central part. For certain observables,
this {}``background'' contribution spoils the {}``signal'', and
to get a detailed understanding of RHIC and SPS data, we need to separate
low and high density parts.

\begin{figure}
~

\vspace{-0.8cm}
\includegraphics[%
  scale=0.25,
  angle=270]{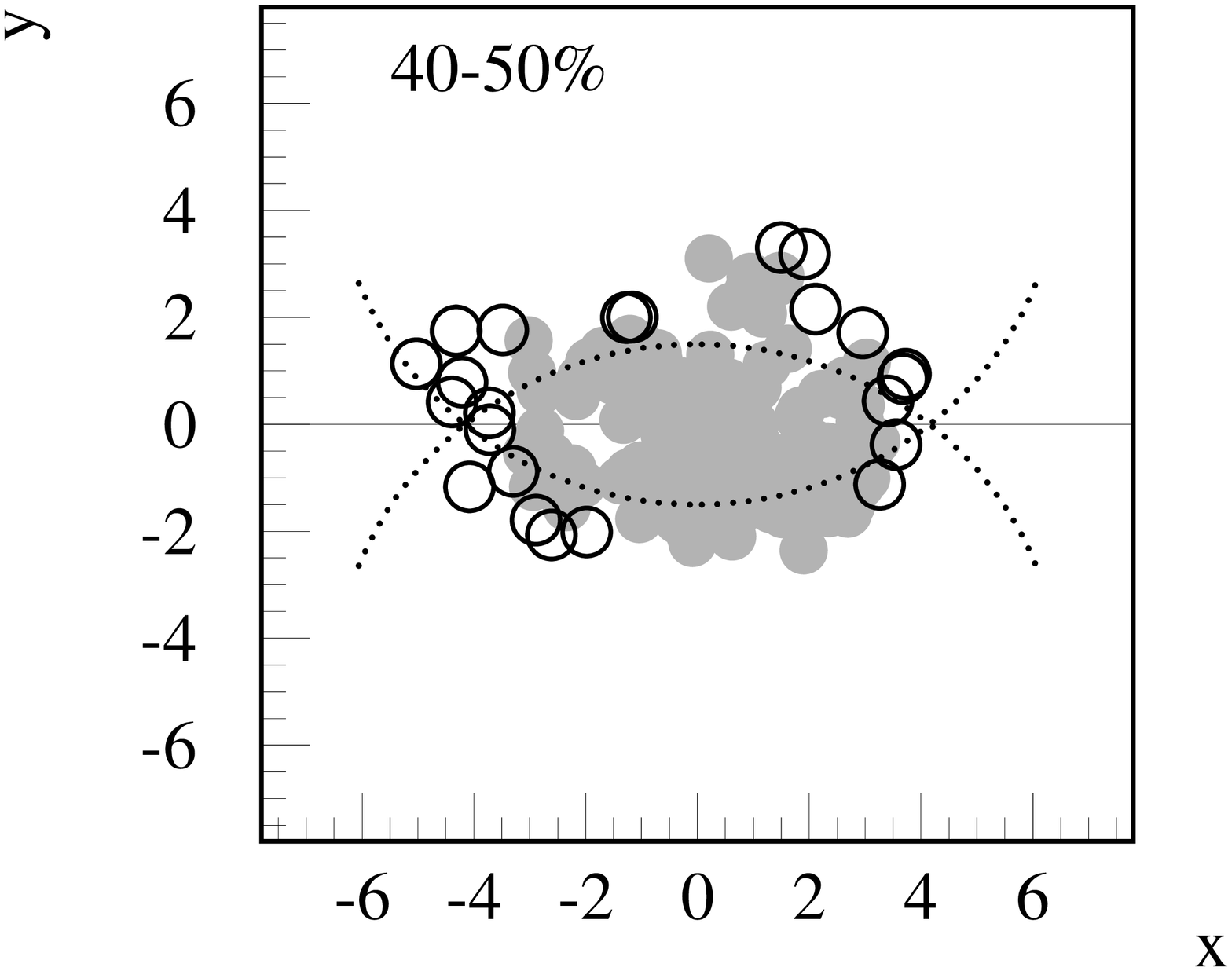}
\vspace{-0.8cm}

\caption{A Monte Carlo realization of a semi-peripheral (40-50\%) AuAu collision
at 200 GeV (cms). We show string segments in the core (full gray circles)
and  the corona (open circles). The big circles are put in just to
guide the eye: they represent the two nuclei in hard sphere approximation.We
consider a projection of segments within $z=\pm0.4\,\mathrm{fm}$
to the transverse plane (x,y).\label{cap:geom1}}
\end{figure}

In order to get quantitative results, we need a simulation tool, and
here we take EPOS \cite{epos}, which has proven to work very well
for pp and dAu collisions at RHIC. EPOS is a parton model, so in case
of a AuAu collision there are many binary interactions, each one represented
by a parton ladder. Such a ladder may be considered as a longitudinal
color field, conveniently treated as a relativistic string. The strings
decay via the production of quark-antiquark pairs, creating in this
way string fragments -- which are usually identified with hadrons.
Here, we modify the procedure: we have a look at the situation at
an early proper time $\tau_{0}$, long before the hadrons are formed:
we distinguish between string segments in dense areas (more than $\rho_{0}$
segments per unit area in given transverse slices), from those in
low density areas. We refer to high density areas as core, and to
low density areas as corona. In figure. \ref{cap:geom1}, we show
an example (randomly chosen) of a semi-peripheral (40-50\%) AuAu collisions
at 200 GeV (cms), simulated with EPOS.

There is always a contribution from the low density area, but much
more importantly, as discussed later, the importance of this contribution
depends strongly on particle type and transverse momentum. For central
collisions, the low density contribution is obviously less important,
for more peripheral collisions this contribution will even dominate.

We adopt the following strategy: the low density part will be treated
using the usual EPOS particle production which has proven to be very
successful in pp and dAu scattering (the peripheral interactions are
essentially pp or pA scatterings). For the high density part, we simply
try to parameterize particle production, in the most simple way possible
(it is not at all our aim to provide a microscopic description of
this part). 

In practice, we consider transverse slices characterized by some range
in $\eta=0.5\ln(t+z)/(t-z)$. String segments in such a slice move
with rapidities very close to $\eta$. We subdivide a given slice
into elementary cells, count the number of string segments per cell,
and determine such for each cell whether it belongs to the core or
the corona (bigger or smaller than the critical density $\rho_{0}$).
Connected cells (closest neighbors) in a given slice are considered
to be clusters, whose energy and flavor content are completely determined
by the corresponding string segments. Clusters are then considered
to be collectively expanding: Bjorken-like in longitudinal direction
with in addition some transverse expansion. We assume particles to
freeze out at some given energy density $\varepsilon_{\mathrm{FO}}$,
having acquired at that moment a collective radial flow. The latter
one is characterized by a linear radial rapidity profile from inside
to outside with maximal radial rapidity $y_{\mathrm{rad}}$. In addition,
we impose an azimuthal asymmetry, being proportional to the initial
spatial eccentricity $\epsilon=\left\langle y^{2}-x^{2}\right\rangle /\left\langle y^{2}+x^{2}\right\rangle $,
with a proportionality factor $f_{\mathrm{ecc}}$. By imposing radial
flow, we have to rescale the cluster mass $M$ as\[
M\to M\,\times\,0.5\, y_{\mathrm{rad}}^{2}/(y_{\mathrm{rad}}\sinh y_{\mathrm{rad}}-\cosh y_{\mathrm{rad}}+1),\]
in order to conserve energy. Hadronization then occurs according to
covariant phase space, which means that the probability $dP$ of a
given final state of n hadrons is given as\[
\hspace{-0.8cm}\prod_{\mathrm{species}\,\alpha}\!\!\!{\frac{1}{n_{\alpha}!}}\prod_{i=1}^{n}\,\frac{d^{3}p_{i}\, g_{i}\, s_{i}}{(2\pi\hbar)^{3}2E_{i}}\,\frac{M}{\varepsilon_{\mathrm{FO}}}\,\delta(M-\Sigma E_{i})\,\delta(\Sigma\vec{p}_{i})\,\delta_{f,\Sigma f_{i}},\]
with $p_{i}=(E_{i},\vec{p_{i}})$ being the four-momentum of the i-th
hadron, $g_{i}$ its degeneracy, and $f_{i}$ its quark flavor content
($u-\bar{u},$$d-\bar{d}$...). The number $n_{\alpha}$ counts the
number of hadrons of species $\alpha$. The term $M/\varepsilon_{\mathrm{FO}}$
is the cluster proper volume. We use a factor $s_{i}=\gamma_{s}\,^{\pm1}$
for each strange particle (sign plus for a baryon, sign minus for
a meson), with $\gamma_{s}$ being a parameter. We believe that $s_{i}$
mimics final state rescattering, but for the moment we can only say
that this factor being different from unity improves the fit of the
data considerably. The whole procedure perfectly conserves energy,
momentum, and flavors (microcanonical procedure).

So the core definition and its hadronization are parameterized in
terms of few global parameters (in brackets the values): the core
formation time $\tau_{0}$ (1$\,$fm), the core formation density
$\rho_{0}$ (2/fm$^{2}$), the freeze out energy density $\varepsilon_{\mathrm{FO}}$(0.22$\,$GeV/fm$^{3}$),
the maximum radial flow $y_{\mathrm{rad}}$ (0.75+0.20$\log(\sqrt{s}/200\mathrm{\, GeV})$),
the eccentricity coefficient $f_{\mathrm{ecc}}$ (0.45), and the factor
$\gamma_{s}$ (1.3). At RHIC energies, the final results are insensitive
to variations of $\tau_{0}$: even changes as big as a factor of 2
do not affect the results. This is a nice feature, indicating that
the very details of the initial state do not matter so much. We call
these parameters {}``global'', since they account for all observables
at all possible different centralities and all energies. In the following,
we are going to discuss results, all obtained with the above set of
parameters.

Our hadronization of the core part is certainly motivated by the remarkable
success of statistical hadronization models \cite{stat} and blast-wave
fits \cite{blast,blast-appl}. We use covariant statistical hadronization,
whereas usual the non-covariant procedure is employed, but the difference
is minor. We also impose a collective flow, with an assumed flow profile,
as in the blast wave fit. So the general ideas are the same. However,
a really new aspect is the possibility of making a {}``global fit'',
considering all energies, centralities, and colliding systems with
one set of parameters. In the above-mentioned models one has a set
of fit parameters for each of these possibilities. An important new
aspect is also the separation of a (collectively behaving) core and
a corona contribution, which seems to be very important for understanding
the centrality dependence of hadron yields. Finally, our statistical
hadronization is based on initial energy densities provided by a parton
model (EPOS), which works well for pp and dAu scattering. This fixes
the overall multiplicity already within 10\%, flow and freeze out
condition have only a minor effect on this quantity. 

\begin{figure}
\begin{center}\includegraphics[%
  scale=0.3,
  angle=270]{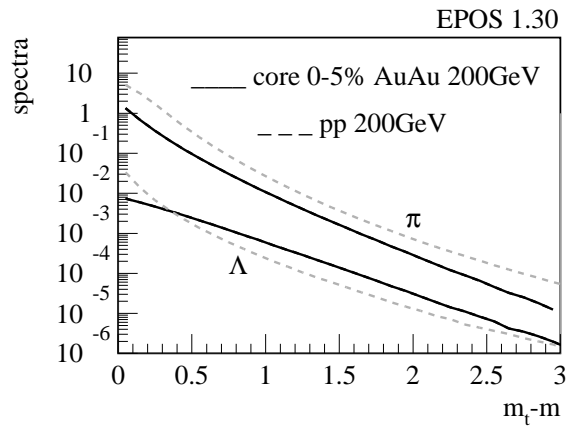}\end{center}
\vspace{-0.95cm}

\caption{Invariant yields $1/2\pi m_{t}\, dn/dydm_{t}$ of pions and lambdas,
for the core contribution corresponding to a central (0-5\%) AuAu
collision (full lines) and proton-proton scattering (dashed lines).
The core spectra are divided by the number of binary collisions.\label{cap:core-pp}}
\end{figure}
\begin{figure}
\begin{center}\hspace*{-1.5cm}\includegraphics[%
  scale=0.3,
  angle=270]{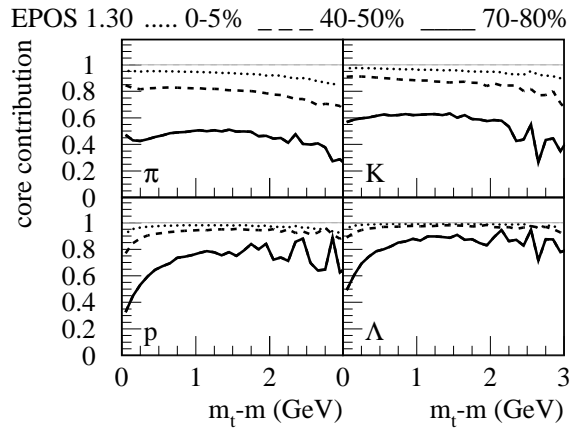}\end{center}
\vspace{-0.95cm}

\caption{The relative contribution of the core (core/(core+corona)) as a function
of the transverse mass for different hadrons ($\pi$, $K$, $p$,
$\Lambda$) at different centralities. \label{cap:core-corona}}
\end{figure}

All the discussion of heavy ion data will be based on the interplay
between core and corona contributions. To get some feeling, we first
compare in fig. \ref{cap:core-pp} the $m_{t}$ spectra of pions and
lambdas from the core in central (0-5\%) AuAu collisions with the
corresponding spectra in pp scattering (which is qualitatively very
similar to the corona contribution). The core spectra are divided
by the number of binary collisions. We observe several remarkable
features: the shapes of the pion and lambda curves in pp are not so
different, whereas there is much more species dependence in the core
spectra, since the heavier particles acquire large transverse momenta
due to the flow effect. One observes furthermore that the yields for
the two spectra in pp are much wider spread than the ones from the
core; in particular, pion production is suppressed in the core hadronization
compared to pp, whereas lambda production is favored. All this is
quite trivial, but several {}``mysteries'' discussed in the literature
(and to be discussed later in this paper) are just due to this.

In fig. \ref{cap:core-corona}, we plot the relative contribution
of the core (relative to the complete spectrum, core + corona) as
a function of $m_{t}-m$, for different particle species. For central
collisions, the core contribution dominates largely, whereas for semi-central
collisions (40-50\%) and even more for peripheral collisions the core
contribution decreases, giving more and more space for the corona
part. Apart of these general statements, the precise $m_{t}$ dependence
of the relative weight of core versus corona depends on the particle
type.

\begin{figure}
\begin{center}\includegraphics[%
  scale=0.3,
  angle=270]{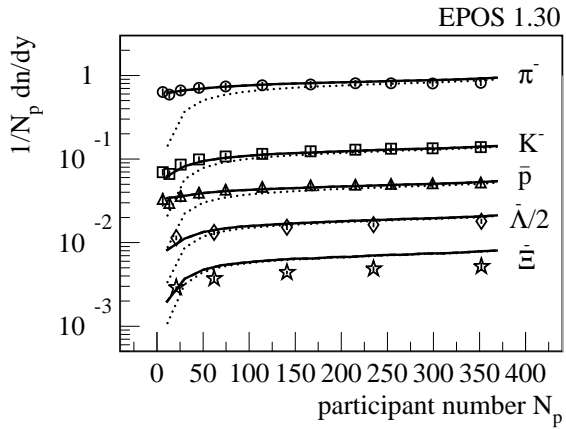}\end{center}
\vspace{-0.95cm}

\caption{Rapidity density dn/dy per participant as a function of the number
of participants ($N_{p}$) in Au+Au collisions at 200 GeV (RHIC) for
$\pi^{-}$, $K^{-}$, $\bar{p}$, $\bar{\Lambda}$, $\bar{\Xi}^{+}$.
We show data (points) \cite{phenix,star-lda} together with the full
calculation (full lines) and just the core part (dotted lines).\label{cap:centrality}}
\end{figure}
\begin{figure}
\begin{center}\includegraphics[%
  scale=0.3,
  angle=270]{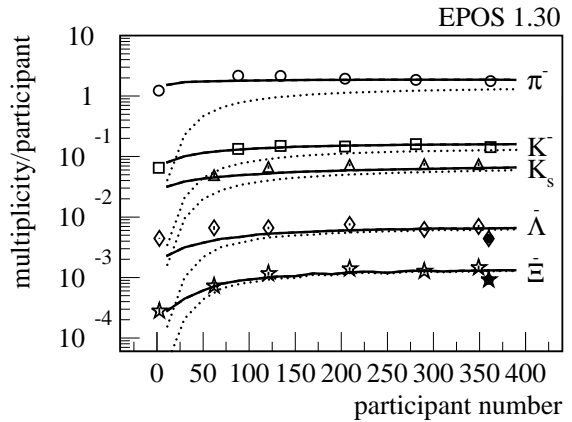}\end{center}
\vspace{-0.95cm}

\caption{Multiplicity per participant as a function of the number of participants
($N_{p}$) in Pb+Pb collisions at 17.3 GeV (SPS) for $\pi^{-}$, $K^{-}$,
$K_{s}$, $\bar{\Lambda}$, $\bar{\Xi}^{+}$. We show data (points)
\cite{na49,na57,na49-alda,na49-xi} together with the full calculation
(full lines) and just the core part (dotted lines).\label{cap:centralitysps}}
\end{figure}

We are now ready to investigate data. In fig. \ref{cap:centrality},
we plot the centrality dependence of the particle yield per participant
(per unit of rapidity) in Au+Au collisions at 200 GeV (RHIC), for
$\pi^{+}$, $K^{+}$, $p$, $\bar{\Lambda}$, $\bar{\Xi}^{+}$: we
show data \cite{phenix,star-lda} together with the full calculation
(quite close to the data), but also indicating the core contribution.
In fig. \ref{cap:centralitysps}, we show the corresponding results
for Pb+Pb collisions at 17.3 GeV (SPS). Concerning the SPS results,
we consider $dn/dy/N_{p}$ in case of $K_{s}$, $\bar{\Lambda}$,
and $\bar{\Xi}^{+}$, whereas we have $4\pi$ multiplicities per participant
in case of $\pi^{-}$ and $K^{-}$(for simulations and data). Whereas
central collisions are always clearly core dominated, the core contributes
less and less with decreasing centrality. The difference between solid
and dotted curves (in other words: the importance of the corona contribution)
is bigger at the SPS compared to RHIC, and it is bigger for light
particles compared to heavy ones. For example there is a big corona
contribution for pions and a very small one for $\bar{\Xi}$ particles.
Also the strength of the centrality dependence depends on the hadron
type: for example $\bar{\Xi}^{+}$particles show a stronger centrality
dependence than pions. It seems that the centrality dependence is
essentially determined by relative importance of the corona contribution:
the less the corona contributes, the more the yield varies with centrality.

\begin{figure}
\begin{center}\includegraphics[%
  scale=0.3,
  angle=270]{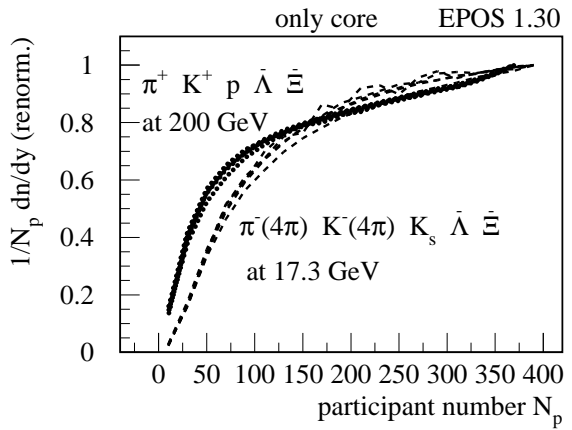}\end{center}
\vspace{-0.95cm}

\caption{Multiplicity per participant as a function of $N_{p}$ for only the
core part. We show results for $\pi^{-}$, $K^{-}$, $\bar{p}$, $\bar{\Lambda}$,
$\bar{\Xi}^{+}$ in Au+Au collisions at 200 GeV (dotted lines), and
for $\pi^{-}$, $K^{-}$, $K_{s}$, $\bar{\Lambda}$, $\bar{\Xi}^{+}$
in Pb+Pb collisions at 17.3 GeV (dashed lines). \label{cap:ratios}}
\end{figure}

To further investigate the connection between relative corona weight
and centrality dependence, we plot in fig. \ref{cap:ratios} the centrality
dependence of multiplicities per participant for different hadrons,
at 200 GeV (RHIC) and 17.3 GeV (SPS), for the core contribution. We
observe two universal curves, one per energy. So for a given energy,
the core contributions for all the different hadrons show the same
centrality dependence. This proves that the different centrality dependencies
for the different hadron species are simply due to different core-corona
weights. For example the fact that $\bar{\Xi}$ particles show a stronger
centrality dependence than pions is simply due to the fact that the
former ones have less corona admixture than the latter ones.

\begin{figure}
\begin{center}\includegraphics[%
  scale=0.3,
  angle=270]{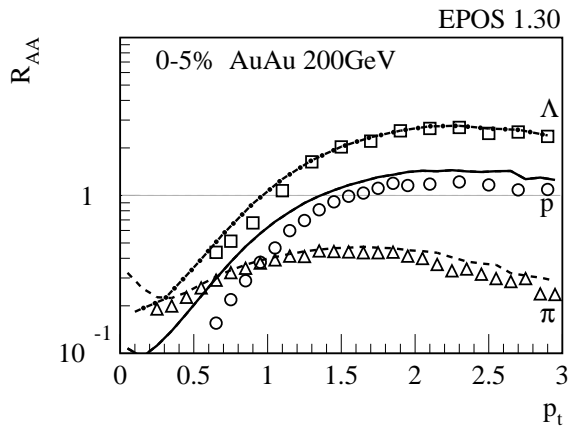}\end{center}
\vspace{-0.95cm}

\caption{Nuclear modification factors in central AuAu collisions at 200 GeV.
Lines are full calculations, symbols represent data \cite{phenix,star-lda}.
We show results for pions (dashed line; triangles), protons (full
line; circles), and lambdas (dashed-dotted line; squares). \label{cap:nmf}}
\end{figure}

Lets us come to $p_{t}$ spectra. We checked all available $p_{t}$
data ($\pi^{+}$, $K^{+}$, $p$, $\bar{\Lambda}$, $\bar{\Xi}^{+}$,
for $p_{t}\leq5\,\mathrm{GeV}$), and our combined approach (core
+ corona) describes all the data within 20\%. Lacking space, we just
discuss a (typical) example: the nuclear modification factor (AA/pp/number
of collisions), for $\pi^{+}$, $p$, $\bar{\Lambda}$ in central
AuAu collisions at 200 GeV, see fig. \ref{cap:nmf}. For understanding
these curves, we simply have a look at fig. \ref{cap:core-pp}, where
we compare the core contributions from AuAu (divided by the number
of binary collisions) with pp. Since for very central collisions the
core dominates largely, the ratio of core to pp (the solid lines divided
by the dotted ones in fig. \ref{cap:core-pp}) corresponds to the
nuclear modification factor. We discussed already earlier the very
different behavior of the%
 core spectra (flow plus phase space decay) compared to the pp spectra
(string decay): pions are suppressed, whereas heavier particles like
lambdas are favored. Or better to say it the other way round: the
production of baryons compared to mesons is much more suppressed in
string decays than in statistical hadronization. This is why the nuclear
modification factor for lambdas is different from the one for pions.
So what we observe here is nothing but the very different behavior
of statistical hadronization (plus flow) on one hand, and string fragmentation
on the other hand. This completely statistical behavior indicates
that the low $p_{t}$ partons get completely absorbed in the core
matter. 

The $R_{cp}$ modification factors (central over peripheral) are much
less extreme than $R_{AA}$, since peripheral AuAu collisions are
a mixture of core and corona (the latter one being pp-like), so a
big part of the effect seen in $R_{AA}$ is simply washed out. 

To summarize: we have discussed the importance of separating core
and corona contributions in ultra-relativistic heavy ion collisions.
The core-corona separation is realized based on the determination
of string densities at an early time. Particle production from the
corona is done as in proton-proton scattering, whereas the core hadronization
is parameterized in a very simple way, imposing radial flow. The corona
contribution is quite small (but not negligible) for central collisions,
but gets increasingly important with decreasing centrality. The core
shows a very simple centrality dependence: it is the same for all
hadron species, at a given bombarding energy. The fact that the centrality
dependence of the total hadron yield is strongly species dependent,
is simply due to the fact that the relative corona contribution depends
on the hadron type.

\end{document}